\documentclass[preprint,aps,nofootinbib]{revtex4}
\usepackage{}
\usepackage{graphicx}%Include figure files
\usepackage{amsmath}
\usepackage{amsfonts}
\usepackage{mathrsfs}
\usepackage{amssymb}
\usepackage{color}%
\usepackage{dcolumn}% Align table columns on decimal point
\providecommand{\U}[1]{\protect\rule{.1in}{.1in}}

\newcommand{\f}{\begin{equation}}
\newcommand{\ff}{\end{equation}}
\newcommand{\fa}{\begin{eqnarray}}
\newcommand{\ffa}{\end{eqnarray}}

\begin{document}

\title{The quantum effect on Friedmann equation in FRW universe}

\author{Wei Zhang $^{1}$}
\email{alqphh@163.com}
\author{Xiao-Mei Kuang $^{2}$}
\email{xmeikuang@gmail.com}

\affiliation{$^1$Department of General Studies, Nanchang
Institute of Science $\&$ Technology, Nanchang 330108, China \\$^2$Center for Gravitation and Cosmology, College
of Physical Science and Technology, Yangzhou University,
Yangzhou 225009, China}

\begin{abstract}
We study the modified Friedmann equation in the
Friedmann-Robertson-Walker universe with quantum effect. Our modified results mainly stem from
the new entropy-area relation and the novel idea of T. Padmanabhan, who considers the cosmic
space to be emerged from the cosmic time progresses, so that the expansion
rate of the universe is determined by  the difference of degrees of freedom between the
 holographic surface and the bulk inside. We also discuss the possibility of having bounce cosmological solution from the modified Friedmann equation in spatially flat geometry.
\end{abstract}
 \maketitle

\section {Introduction}
In the 1970s, the thermodynamic property of black holes has been
proposed\cite{Bekenstein:1973ur,Bardeen:1973gs,Hawking:1975iha},
and it reveals that the gravitational dynamics is entwined with thermodynamics.
Inspired by Bekenstein's entropy-area theorem\cite{Bekenstein:1973ur}, Bardeen, Carter and Hawking put
forward the four thermodynamical laws of black hole systems\cite{Bardeen:1973gs}. In 1995, Jacobson
considered the Einstein's field equation as an equation of state. Afterwards, he reproduced the Einstein's field equation by demanding that the fundamental  relation $\delta Q=T d S$ holds for all local Rindler causal horizons through each spacetime point, and
$\delta Q$ and $T$ are treated as the energy flux and Unruh temperature, respectively,
felt by an accelerated observer inside the horizon\cite{Jacobson:1995ab}. In 2010,
Verlinde defined gravity as an entropic
force due to the changes of the information related to  the positions of the materials, and the space is emergent based on the holographic principle in his discussions\cite{Verlinde:2010hp}. Moreover, Verlinde's proposal has been applied to reproduce the Friedmann equation
into brane cosmology\cite{Ling:2010zc} and Friedmann-Robertson-Walker(FRW) universe\cite{Cai:2010hk}, respectively.

On the other hand, it was addressed in \cite{Bojowald:2006da,Ashtekar:2006rx,Ashtekar:2006wn,Ashtekar:2006es,Ashtekar:2008gn} that the Friedmann equation can be modified by a bounce solution of the universe as
\begin{equation}\label{eq1}
H^2=\frac{8\pi G}{3}\rho(1-\frac{\rho}{\rho_c})
\end{equation}
in loop quantum cosmology(LQC). Then the authors of \cite{Cai:2008ys} attempted to derive the Friedmann equation by
borrowing the  clausius relation, i.e., $\delta Q=TdS$ and an entropy-area
relation with quantum correction, however, they failed to reproduce the same modified Friedmann equation as that in LQC with bounce solution. Luckily, the difficulty was overcome by the authors of \cite{Ling:2009wj}, where they proposed a modified dispersion relation at quantum phenomenological level and then obtained the modified Friedmann equation for a bounce solution to the flat FRW universe in LQC. It was found that the role of their modified dispersion
relation is to  explicitly modify the clausius relation. This proposal has also been extended into the spatially curved cases
and the corresponding modified entropy-area relations have been derived\cite{Pan:2015tza}.

However, the above studies only involved in the gravity as an emergent phenomenon
rather than the spacetime itself as an emergent structure. This
situation has been improved by Padmanabhan. In details, he proposed in \cite{Padmanabhan:2012ik} that the accelerated expansion of the universe is related with the difference between the surface degrees of freedom ($N_{sur}$) and the bulk degrees of freedom ($N_{bulk}$) in a region of space, i.e., $\Delta V=\Delta t(N_{sur}-N_{bulk})$, where $V$ is the Hubble volume and $t$ is the cosmic time in Planck
units. Moreover, the standard evolution of the universe was also reproduced directly from the proposed relation.
This proposal inspired plenty of related studies and remarkable progress\cite{Yang:2012wn,Tu:2013gna,Ling:2013qoa,Eune:2013ima, Tu:2013fpa,Ai:2013jha,Hashemi:2013oia,Sheykhi:2013tqa,Chang-Young:2013gwa,
Sheykhi:2013ffa,Sepehri:2014jla,Wang:2015cna,Sepehri:2016jfx,Yuan:2016pkz,
Komatsu:2016vof,B:2017kyi,Tu:2017jen}. However, in the framework of Padmanabhan's
conjecture, the study of quantum effect is missing. So, it is interesting to introduce the quantum effect to Padmanabhan's conjecture
and study the related cosmology.

Thus, in this paper, we will introduce the modified dispersion relation in the framework of Padmanabhan's conjecture and derive the modified Friedmann equation. Then we will analyze whether the quantum effect in Padmanabhan's conjecture
will bring in the modified Friedmann equation \eqref{eq1} with bounce solution. It is notable that starting from the clausius relation to the apparent horizon along with the modified dispersion relation,  one can easily get the modified Friedmann equation with bounce solution to the FRW universe\cite{Ling:2009wj,Pan:2015tza}, but the answer is not direct in Padmanabhan's conjecture in the emergent universe. Our study will give an insight to the answer.

Our paper is organized as follows. In next section we briefly review
Padmanabhan's idea that the cosmic space is emergent as cosmic time progresses,
and give the standard Friedmann equation governing the dynamical evolution
of the FRW universe. Then in section \ref{sec:3}, we will analyze the modified Friedmann
equation from Padmanabhan's conjecture based on modified entropy-area relation.
Finally, we will give our summary and discussions in section \ref{sec:4}. In this paper,
we use the natural units with $c=\hbar =k_B=1$.

\section {The emergence of cosmic space of the FRW universe}\label{sec:2}
In this section, we will give a brief review on the process of obtaining standard Friedmann equation in the emergent universe,
which was addressed by Padmanabhan in \cite{Padmanabhan:2012ik}. The main idea is that the expansion of the universe, (or the emergence of space) tends to fulfill the holographic equipartition condition, which is stated that the number of degrees of freedom ($N_{bulk}$)  inside the Hubble volume is equal to the number of degrees of freedom  ($N_{sur}$) on the
spherical surface of Hubble radius, i.e., $N_{bulk}=N_{sur}$. So in our asymptotically de Sitter universe, the natural
law governing the emergence of space in an infinitesimal interval $dt$ is
\begin{equation}
 \frac{dV}{dt}=G(N_{sur}-N_{bulk}), \label{faf}
\end{equation}
where $V=\frac{4\pi}{3H^3}$ is the Hubble volume and $t$ is the cosmic time.

For a spatially flat FRW universe with Hubble constant $H$ and apparent horizon $r_A=1/H$, we have
\begin{equation}
    N_{sur}=4S=\frac{4\pi}{G H^2}, \label{faa}
\end{equation}
where $S=\frac{A}{4G}=\pi/G  H^2$ is the entropy of the apparent horizon, and
\begin{equation}
 N_{bulk}=\frac{2|E|}{T}=-\frac{2(\rho +3p)V}{T},\label{fab}
\end{equation}
where in the second equality, we recall the horizon temperature $T=\frac{H}{2\pi}$ and Komar energy
$|E|=-(\rho +3p)V$ for accelerating part with dark energy\footnote{It is notable that in \cite{Padmanabhan:2012ik}, the author discussed the contributions of the matter with $|E|=(\rho +3p)V$ in the bulk degrees of freedom and the derivation of Friedmann equation was unaffected.}. Subsequently, one can reduce \eqref{faf} into
\begin{equation}
    \frac{\ddot a}{a}=-\frac{4\pi G }{3}(\rho +3p), \label{fag}
\end{equation}
which is the standard dynamical Friedmann equation of flat FRW universe
in general relativity. Furthermore, recalling the continuity equation
\begin{equation}
\dot\rho+3H(\rho+3p)=0, \label{fah}
\end{equation}
and integrating (\ref{fag}) gives us the standard Friedmann equation
\begin{equation}
    H^2=\frac{8\pi G \rho}{3}. \label{fai}
\end{equation}
Note that in \cite{Cai:2008ys}  the integration result is $H^2+\frac{k}{a^2}=\frac{8\pi G \rho}{3}$ with general geometry, where the authors interpreted the integration constant $k$  as the spatial curvature of the FRW universe.

\section {The quantum effect on Friedmann equation in cosmic space of the FRW universe}\label{sec:3}
In this section, we will apply the proposal described in last section to study the quantum effect on the Friedmann equation. We only consider the quantum effect at the phenomenological level and borrow the modified dispersion relation(MDR)\cite{Ling:2009wj}
\begin{equation}\label{dispersion}
\frac{\sin(\eta l_p E)}{\eta l_p}=\sqrt{p^2+m^2}.
\end{equation}
Here $p$ and $E$ are the momentum and energy of a particle with mass $m$, respectively. The Planck length is
$l_p=\sqrt{8\pi G}=1/M_p$ where $M_p$ is the Planck mass. $\eta$ is a dimensionless parameter and $\eta\to 0$ goes to the standard dispersion relation $E^2=p^2+m^2$.

With the use of thermodynamical description on the apparent horizon, the authors of \cite{Ling:2009wj} derived the modified Friedmann equation of a spatially
flat universe from the MDR \eqref{dispersion}. Later, the extended study in general FRW universe with $k=0,\pm1$ was presented in \cite{Pan:2015tza}.

Here, we will derive the modified Friedmann equation by following the steps of emergent cosmic space shown in last section.
According to the study in \cite{Pan:2015tza}, the MDR \eqref{dispersion} modified the entropy for the first energy branch as
\begin{equation}
S_M=\frac{A}{4G}\sqrt{1-\frac{4\pi \eta^2 l_p^2}{A}}+4\pi \eta^2 l_p^2\ln[\sqrt{\frac{A}{4\pi \eta^2 l_p^2}}+\sqrt{\frac{A}{4\pi \eta^2 l_p^2}-1}],
\end{equation}
where $A=4\pi r_A^2=4\pi/(H^2+k/a^2)$ is the area of the apparent horizon at the classical level. To proceed, we define an effective apparent horizon area with the quantum effect
\begin{eqnarray}
\label{fff} {\widetilde A}=4GS_{M}&=&A\sqrt{1-\frac{4\pi \eta^2 l_p^2}{A}}+4\pi \eta^2 l_p^2\ln[\sqrt{\frac{A}{4\pi \eta^2 l_p^2}}+\sqrt{\frac{A}{4\pi \eta^2 l_p^2}-1}]\nonumber\\
&=&4\pi r_A^2\sqrt{1-\frac{\eta^2 l_p^2}{r_A^2}}+4\pi \eta^2 l_p^2\ln[\frac{r_A}{\eta l_p}+\sqrt{\frac{r_A^2}{\eta^2 l_p^2}-1}]. \label{ffg}
\end{eqnarray}
Note that when $\eta\to 0$, $\widetilde A$ is equal to $A$ and recovers the usual result.

Moreover, the volume ($V$) and the area ($A$) of the apparent horizon of an $n$-sphere with radius $r_A$ satisfies
\cite{Cai:2012ip}
\begin{equation}
    \frac{dV}{dA}=\frac{r_A}{n-1}. \label{a}
\end{equation}
Then one can think that the change of the effective volume mainly stems from the change of the effective area, so that
we have the time evolution of the effective volume of the FRW universe\cite{Cai:2012ip}
\begin{equation}
    \frac{d\widetilde V}{dt}=\frac{r_A}{2} \frac{d\widetilde A}{dt}=\frac{6\pi r_A^3\dot{r}_A}{\eta l_p\sqrt{\frac{r_A^2}{\eta^2 l_p^2}-1}}-\frac{2\pi \eta l_p r_A\dot{r}_A}{\sqrt{\frac{r_A^2}{\eta^2 l_p^2}-1}}-2\pi \eta l_p r_A\dot{r}_A\sqrt{\frac{r_A^2}{\eta^2 l_p^2}-1}, \label{ffj}
\end{equation}
from which we can obtain the effective volume
\begin{equation}
    \widetilde V=\frac{4\pi \eta^3 l_p^3}{3}(2+\frac{r_A^2}{\eta^2 l_p^2})\sqrt{\frac{r_A^2}{\eta^2 l_p^2}-1}. \label{ffk}
\end{equation}
Also, when $\eta\to 0$, $\widetilde V=\frac{4\pi r_A^3}{3}$ is the usual Hubble volume.

We move on to calculate $N_{bulk}$ in the bulk and $N_{sur}$ in the boundary.  Considering the Hawking temperature $T=\frac{1}{2\pi r_A}$ and $E=-(\rho+3p)\widetilde V$ with dark energy in the bulk, we obtain
\begin{equation}
    N_{bulk}=\frac{2E}{T}=-\frac{16\pi^2(\rho+3p)\eta^3 l_p^3 r_A}{3}(2+\frac{r_A^2}{\eta^2 l_p^2})\sqrt{\frac{r_A^2}{\eta^2 l_p^2}-1}. \label{ffl}
\end{equation}
The statistical physics has shown that $N_{sur}$ can be calculated from the
entropy\cite{Tu:2013gna}
\begin{equation}
  N_{sur}=4S_M=\frac{4\pi r_A^2}{G}\sqrt{1-\frac{\eta^2 l_p^2}{r_A^2}}+\frac{4\pi \eta^2 l_p^2}{G}\ln[\frac{r_A}{\eta l_p}+\sqrt{\frac{r_A^2}{\eta^2 l_p^2}-1}]. \label{ffn}
\end{equation}
Substituting \eqref{ffj}, \eqref{ffl} and \eqref{ffn} into \eqref{faf}, we get
\begin{eqnarray}
  &&  \frac{6\pi r_A^3\dot{r}_A}{\eta l_p\sqrt{\frac{r_A^2}{\eta^2 l_p^2}-1}}-\frac{2\pi \eta l_p r_A\dot{r}_A}{\sqrt{\frac{r_A^2}{\eta^2 l_p^2}-1}}-2\pi \eta l_p r_A\dot{r}_A\sqrt{\frac{r_A^2}{\eta^2 l_p^2}-1}= 4\pi r_A^2\sqrt{1-\frac{\eta^2 l_p^2}{r_A^2}}\nonumber\\
   && +4\pi \eta^2 l_p^2\ln[\frac{r_A}{\eta l_p}+\sqrt{\frac{r_A^2}{\eta^2 l_p^2}-1}]+\frac{16\pi^2G(\rho+3p)\eta^3 l_p^3 r_A}{3}(2+\frac{r_A^2}{\eta^2 l_p^2})\sqrt{\frac{r_A^2}{\eta^2 l_p^2}-1}.\label{ffp}
\end{eqnarray}

The expression above looks very complicated, however, with $k=0$, we have $\dot{r}_A=1-\frac{\ddot a}{a}r_A^2$,
so that equation \eqref{ffp} can be reduced into\footnote{With $k=\pm1$, we have $\dot{r}_A=1-\frac{H\dot{H}-\frac{k\dot{a}}{a^3}+\sqrt{(H^2+\frac{k}{a^2})^3}}{\sqrt{(H^2+\frac{k}{a^2})}}r_A^2$
which makes it difficult to simplify equation \eqref{ffp}. We hope to solve this problem in near future.}
\begin{eqnarray}
    \frac{\ddot a}{a}&=&-\frac{4\pi G(\rho+3p)}{3}(1+\frac{\eta^2 l_p^2}{r_A^2}-\frac{2\eta^4 l_p^4}{r_A^4})+\frac{\eta^2 l_p^2[2+\ln(\frac{\eta^2 l_p^2}{4r_A^2})]}{2r_A^4} \label{ffs}\nonumber\\
     &\simeq& -\frac{4\pi G(\rho+3p)}{3}(1+\frac{\eta^2 l_p^2}{r_A^2})+\frac{\eta^2 l_p^2}{r_A^4}, \label{fft}
\end{eqnarray}
where in the second line, we keep until the second order of $\frac{\eta^2 l_p^2}{r_A^2}$ because it is a small quantity. The form \eqref{fft} is the modified dynamical Friedmann equation for the flat FRW universe, which reduces to the standard dynamical
Friedmann equation when $\eta\to 0$.
Further combining the continuity equation \eqref{fah} with \eqref{fft}, we obtain  the other modified Friedmann equation
\begin{equation}
    \frac{\dot a^2}{a^2}=\frac{8\pi G\rho}{3}+\frac{8\pi G\eta^2 l_p^2}{3a^2}\int(\dot a^2\dot\rho+\frac{2\dot a^3\rho}{a})dt+\frac{2\eta^2 l_p^2}{a^2}\int\frac{\dot a^5}{a^3}dt. \label{ffv}
\end{equation}
Here we also set the integration constant to be vanished. Again,
$\eta\to 0$ in \eqref{ffv} reproduces the standard result of flat FRW universe.

In order to analyze whether \eqref{ffv} admits a bounce solution, we define
\begin{equation}
    \rho_c=-\frac{\rho^2}{\frac{\eta^2 l_p^2}{a^2}\int(\dot a^2\dot\rho+\frac{2\dot a^3\rho}{a})dt+\frac{3\eta^2 l_p^2}{4\pi G\rho a^2}\int\frac{\dot a^5}{a^3}dt}, \label{ffr}
\end{equation}
so that equation \eqref{ffv} can be rewritten as equation \eqref{eq1} for bounce solution.
The unsolved integral in $\rho_c$ makes it difficult to give a reliable conclusion, however, we can at least give some discussions. On one hand, when $\rho_c$ in \eqref{ffr} is positive, Eqs.\eqref{fft} and \eqref{ffv} fulfill the bouncing conditions, i.e., $a>0$, $\dot{a}=0$ and $\ddot{a}>0$, then \eqref{ffv} admits a bounce solution. On the other hand, when $\rho_c$  is non-positive, we can not have any bounce solution.

We note that for the second energy branch whose entropy is $-S_M$\cite{Pan:2015tza}, the procedures above are straightforward and the modified Friedmann equation are the same as equations \eqref{fft} and \eqref{ffv}. However, for this energy branch, the effective volume ($\widetilde V$),  and the area ($\widetilde A$)  of the apparent horizon are negative, which are not physical.

\section{Summary and Discussions}\label{sec:4}
In this paper, we studied the quantum effect on the Friedmann equation for the flat FRW universe with the use of Padmanabhan's conjecture in the emergent universe. We obtained the modified Friedmann equations \eqref{fft} and \eqref{ffv} with the quantum correction on the dispersion relation \eqref{dispersion}.  For the closed ($k=1$) and open ($k=-1$) universes, we found it is difficult to simplify the dynamical equation in our process, but we still see the quantum effect on equation \eqref{ffp} which supposes to be the modified Friedmann equation. We also argued the condition under which the modified Friedmann equation \eqref{ffv} admits a bounce solution in the flat universe.

It is worthy to point out that in our paper,  the modified Friedmann equation was only obtained in the flat FRW universe with $k=0$, which may imply that the key equation \eqref{faf} is not the basic equation of the emergent universe and it may have to be corrected at the quantum level. This is an interesting point we will study in the near future.

\begin{acknowledgments}
We appreciate Yi Ling and Wen-Jian Pan for helpful discussion. This work is supported by the Natural Science
Foundation of China under Grant No. 11705161 and  Natural Science Foundation of Jiangsu Province under Grant No. BK20170481.
\end{acknowledgments}


\begin{thebibliography}{99}%{References}
%\cite{Bekenstein:1973ur}
\bibitem{Bekenstein:1973ur}
  J.~D.~Bekenstein,
  ``Black holes and entropy,''
  Phys.\ Rev.\ D {\bf 7}, 2333 (1973).
  %%CITATION = doi:10.1103/PhysRevD.7.2333;%%
  %3723 citations counted in INSPIRE as of 23 Sep 2017

%\cite{Bardeen:1973gs}
\bibitem{Bardeen:1973gs}
  J.~M.~Bardeen, B.~Carter and S.~W.~Hawking,
  ``The Four laws of black hole mechanics,''
  Commun.\ Math.\ Phys.\  {\bf 31}, 161 (1973).
  %%CITATION = doi:10.1007/BF01645742;%%
  %1622 citations counted in INSPIRE as of 23 Sep 2017

%%\cite{Hawking:1975iha}
\bibitem{Hawking:1975iha}
  S.~W.~Hawking,
  ``Particle Creation by Black Holes,''
  Commun.\ Math.\ Phys.\  {\bf 43}, 199 (1975).
  %%CITATION = INSPIRE-1365034;%%

%\cite{Jacobson:1995ab}
\bibitem{Jacobson:1995ab}
  T.~Jacobson,
  ``Thermodynamics of space-time: The Einstein equation of state,''
  Phys.\ Rev.\ Lett.\  {\bf 75}, 1260 (1995)
  [gr-qc/9504004].
  %%CITATION = doi:10.1103/PhysRevLett.75.1260;%%
  %1105 citations counted in INSPIRE as of 23 Sep 2017

%\cite{Verlinde:2010hp}
\bibitem{Verlinde:2010hp}
  E.~P.~Verlinde,
  ``On the Origin of Gravity and the Laws of Newton,''
  JHEP {\bf 1104}, 029 (2011)
  [arXiv:1001.0785 [hep-th]].
  %%CITATION = doi:10.1007/JHEP04(2011)029;%%
  %635 citations counted in INSPIRE as of 23 Sep 2017

%\cite{Ling:2010zc}
\bibitem{Ling:2010zc}
  Y.~Ling and J.~P.~Wu,
  ``A note on entropic force and brane cosmology,''
  JCAP {\bf 1008}, 017 (2010)
  [arXiv:1001.5324 [hep-th]].
  %%CITATION = doi:10.1088/1475-7516/2010/08/017;%%
  %65 citations counted in INSPIRE as of 24 Sep 2017

%\cite{Cai:2010hk}
\bibitem{Cai:2010hk}
  R.~G.~Cai, L.~M.~Cao and N.~Ohta,
  ``Friedmann Equations from Entropic Force,''
  Phys.\ Rev.\ D {\bf 81}, 061501 (2010)
  [arXiv:1001.3470 [hep-th]].
  %%CITATION = doi:10.1103/PhysRevD.81.061501;%%
  %138 citations counted in INSPIRE as of 24 Sep 2017

%\cite{Bojowald:2006da}
\bibitem{Bojowald:2006da}
  M.~Bojowald,
  ``Loop quantum cosmology,''
  Living Rev.\ Rel.\  {\bf 8}, 11 (2005)
  [gr-qc/0601085].
  %%CITATION = doi:10.12942/lrr-2005-11;%%
  %405 citations counted in INSPIRE as of 08 Oct 2017

%\cite{Ashtekar:2006rx}
\bibitem{Ashtekar:2006rx}
  A.~Ashtekar, T.~Pawlowski and P.~Singh,
  ``Quantum nature of the big bang,''
  Phys.\ Rev.\ Lett.\  {\bf 96}, 141301 (2006)
  [gr-qc/0602086].
  %%CITATION = doi:10.1103/PhysRevLett.96.141301;%%
  %414 citations counted in INSPIRE as of 08 Oct 2017

%\cite{Ashtekar:2006wn}
\bibitem{Ashtekar:2006wn}
  A.~Ashtekar, T.~Pawlowski and P.~Singh,
  ``Quantum Nature of the Big Bang: Improved dynamics,''
  Phys.\ Rev.\ D {\bf 74}, 084003 (2006)
  [gr-qc/0607039].
  %%CITATION = doi:10.1103/PhysRevD.74.084003;%%
  %635 citations counted in INSPIRE as of 08 Oct 2017

%\cite{Ashtekar:2006es}
\bibitem{Ashtekar:2006es}
  A.~Ashtekar, T.~Pawlowski, P.~Singh and K.~Vandersloot,
  ``Loop quantum cosmology of k=1 FRW models,''
  Phys.\ Rev.\ D {\bf 75}, 024035 (2007)
  [gr-qc/0612104].
  %%CITATION = doi:10.1103/PhysRevD.75.024035;%%
  %239 citations counted in INSPIRE as of 08 Oct 2017

%\cite{Ashtekar:2008gn}
\bibitem{Ashtekar:2008gn}
  A.~Ashtekar and E.~Wilson-Ewing,
  ``The Covariant entropy bound and loop quantum cosmology,''
  Phys.\ Rev.\ D {\bf 78}, 064047 (2008)
  [arXiv:0805.3511 [gr-qc]].
  %%CITATION = doi:10.1103/PhysRevD.78.064047;%%
  %31 citations counted in INSPIRE as of 08 Oct 2017

%\cite{Cai:2008ys}
\bibitem{Cai:2008ys}
  R.~G.~Cai, L.~M.~Cao and Y.~P.~Hu,
  ``Corrected Entropy-Area Relation and Modified Friedmann Equations,''
  JHEP {\bf 0808}, 090 (2008)
  [arXiv:0807.1232 [hep-th]].
  %%CITATION = doi:10.1088/1126-6708/2008/08/090;%%
  %145 citations counted in INSPIRE as of 08 Oct 2017

%\cite{Ling:2009wj}
\bibitem{Ling:2009wj}
  Y.~Ling, W.~J.~Li and J.~P.~Wu,
  ``Bouncing universe from a modified dispersion relation,''
  JCAP {\bf 0911}, 016 (2009)
  [arXiv:0909.4862 [gr-qc]].
  %%CITATION = doi:10.1088/1475-7516/2009/11/016;%%
  %7 citations counted in INSPIRE as of 24 Sep 2017

%\cite{Pan:2015tza}
\bibitem{Pan:2015tza}
  W.~J.~Pan and Y.~C.~Huang,
  ``Bouncing universe with modified dispersion relation,''
  Gen.\ Rel.\ Grav.\  {\bf 48}, no. 11, 144 (2016)
  [arXiv:1508.06475 [hep-th]].
  %%CITATION = doi:10.1007/s10714-016-2138-y;%%

%\cite{Padmanabhan:2012ik}
\bibitem{Padmanabhan:2012ik}
  T.~Padmanabhan,
  ``Emergence and Expansion of Cosmic Space as due to the Quest for Holographic Equipartition,''
  arXiv:1206.4916 [hep-th].
  %%CITATION = ARXIV:1206.4916;%%

%\cite{Yang:2012wn}
\bibitem{Yang:2012wn}
  K.~Yang, Y.~X.~Liu and Y.~Q.~Wang,
  ``Emergence of Cosmic Space and the Generalized Holographic Equipartition,''
  Phys.\ Rev.\ D {\bf 86}, 104013 (2012)
  [arXiv:1207.3515 [hep-th]].
  %%CITATION = doi:10.1103/PhysRevD.86.104013;%%
  %29 citations counted in INSPIRE as of 25 Sep 2017

%\cite{Tu:2013gna}
\bibitem{Tu:2013gna}
  F.~Q.~Tu and Y.~X.~Chen,
  ``Emergence of spaces and the dynamic equations of FRW universes in the $f(R)$ theory and deformed Ho\v{r}ava-Lifshitz theory,''
  JCAP {\bf 1305}, 024 (2013)
  [arXiv:1303.5813 [hep-th]].
  %%CITATION = doi:10.1088/1475-7516/2013/05/024;%%
  %13 citations counted in INSPIRE as of 25 Sep 2017

%\cite{Ling:2013qoa}
\bibitem{Ling:2013qoa}
  Y.~Ling and W.~J.~Pan,
  ``Note on the emergence of cosmic space in modified gravities,''
  Phys.\ Rev.\ D {\bf 88}, 043518 (2013)
  [arXiv:1304.0220 [hep-th]].
  %%CITATION = doi:10.1103/PhysRevD.88.043518;%%
  %18 citations counted in INSPIRE as of 25 Sep 2017

%\cite{Eune:2013ima}
\bibitem{Eune:2013ima}
  M.~Eune and W.~Kim,
  ``Emergent Friedmann equation from the evolution of cosmic space revisited,''
  Phys.\ Rev.\ D {\bf 88}, no. 6, 067303 (2013)
  [arXiv:1305.6688 [gr-qc]].
  %%CITATION = doi:10.1103/PhysRevD.88.067303;%%
  %17 citations counted in INSPIRE as of 25 Sep 2017

%\cite{Tu:2013fpa}
\bibitem{Tu:2013fpa}
  F.~Q.~Tu and Y.~X.~Chen,
  ``Emergence of space and cosmic evolution based on entropic force,''
  Gen.\ Rel.\ Grav.\  {\bf 47}, no. 8, 87 (2015)
  [arXiv:1306.0639 [hep-th]].
  %%CITATION = doi:10.1007/s10714-015-1927-z;%%
  %5 citations counted in INSPIRE as of 25 Sep 2017

%\cite{Ai:2013jha}
\bibitem{Ai:2013jha}
  W.~Y.~Ai, H.~Chen, X.~R.~Hu and J.~B.~Deng,
  ``Emergence of space and the general dynamic equation of the Friedmann-Robertson-Walker universe,''
  Phys.\ Rev.\ D {\bf 88}, no. 8, 084019 (2013)
  [arXiv:1307.2480 [gr-qc]].
  %%CITATION = doi:10.1103/PhysRevD.88.084019;%%
  %9 citations counted in INSPIRE as of 25 Sep 2017

%\cite{Hashemi:2013oia}
\bibitem{Hashemi:2013oia}
  M.~Hashemi, S.~Jalalzadeh and S.~Vasheghani Farahani,
  ``Hawking temperature and the emergent cosmic space,''
  Gen.\ Rel.\ Grav.\  {\bf 47}, no. 4, 53 (2015)
  [arXiv:1308.2383 [gr-qc]].
  %%CITATION = doi:10.1007/s10714-015-1893-5;%%
  %5 citations counted in INSPIRE as of 25 Sep 2017

%\cite{Sheykhi:2013tqa}
\bibitem{Sheykhi:2013tqa}
  A.~Sheykhi, M.~H.~Dehghani and S.~E.~Hosseini,
  ``Friedmann equations in braneworld scenarios from emergence of cosmic space,''
  Phys.\ Lett.\ B {\bf 726}, 23 (2013)
  [arXiv:1308.2668 [hep-th]].
  %%CITATION = doi:10.1016/j.physletb.2013.08.035;%%
  %10 citations counted in INSPIRE as of 25 Sep 2017

%\cite{Chang-Young:2013gwa}
\bibitem{Chang-Young:2013gwa}
  E.~Chang-Young and D.~Lee,
  ``Friedmann equation and the emergence of cosmic space,''
  JHEP {\bf 1404}, 125 (2014)
  [arXiv:1309.3084 [hep-th]].
  %%CITATION = doi:10.1007/JHEP04(2014)125;%%
  %10 citations counted in INSPIRE as of 25 Sep 2017

%\cite{Sheykhi:2013ffa}
\bibitem{Sheykhi:2013ffa}
  A.~Sheykhi, M.~H.~Dehghani and S.~E.~Hosseini,
  ``Emergence of spacetime dynamics in entropy corrected and braneworld models,''
  JCAP {\bf 1304}, 038 (2013)
  [arXiv:1309.5774 [gr-qc]].
  %%CITATION = doi:10.1088/1475-7516/2013/04/038;%%
  %6 citations counted in INSPIRE as of 25 Sep 2017

%\cite{Sepehri:2014jla}
\bibitem{Sepehri:2014jla}
  A.~Sepehri, F.~Rahaman, A.~Pradhan and I.~H.~Sardar,
  ``Emergence and Expansion of Cosmic Space in BIonic system,''
  Phys.\ Lett.\ B {\bf 741}, 92 (2015)
  [arXiv:1501.00428 [gr-qc]].
  %%CITATION = doi:10.1016/j.physletb.2014.12.030;%%
  %22 citations counted in INSPIRE as of 25 Sep 2017

%\cite{Wang:2015cna}
\bibitem{Wang:2015cna}
  Z.~L.~Wang, W.~Y.~Ai, H.~Chen and J.~B.~Deng,
  ``A cosmological model from emergence of space,''
  Phys.\ Rev.\ D {\bf 92}, no. 2, 024051 (2015)
  [arXiv:1504.05426 [gr-qc]].
  %%CITATION = doi:10.1103/PhysRevD.92.024051;%%
  %5 citations counted in INSPIRE as of 25 Sep 2017

%\cite{Sepehri:2016jfx}
\bibitem{Sepehri:2016jfx}
  A.~Sepehri, F.~Rahaman, S.~Capozziello, A.~F.~Ali and A.~Pradhan,
  ``Emergence and oscillation of cosmic space by joining M1-branes,''
  Eur.\ Phys.\ J.\ C {\bf 76}, no. 5, 231 (2016)
  [arXiv:1604.02451 [gr-qc]].
  %%CITATION = doi:10.1140/epjc/s10052-016-4084-y;%%
  %11 citations counted in INSPIRE as of 25 Sep 2017

%\cite{Yuan:2016pkz}
\bibitem{Yuan:2016pkz}
  F.~F.~Yuan and P.~Huang,
  ``Emergent cosmic space in Rastall theory,''
  Class.\ Quant.\ Grav.\  {\bf 34}, no. 7, 077001 (2017)
  [arXiv:1607.04383 [gr-qc]].
  %%CITATION = doi:10.1088/1361-6382/aa61df;%%
  %3 citations counted in INSPIRE as of 25 Sep 2017

%\cite{Komatsu:2016vof}
\bibitem{Komatsu:2016vof}
  N.~Komatsu,
  ``Cosmological model from the holographic equipartition law with a modified R¨¦nyi entropy,''
  Eur.\ Phys.\ J.\ C {\bf 77}, no. 4, 229 (2017)
  [arXiv:1611.04084 [gr-qc]].
  %%CITATION = doi:10.1140/epjc/s10052-017-4800-2;%%
  %1 citations counted in INSPIRE as of 25 Sep 2017

%\cite{B:2017kyi}
\bibitem{B:2017kyi}
  P.~B.~Krishna and T.~K.~Mathew,
  ``Holographic equipartition and the maximization of entropy,''
  Phys.\ Rev.\ D {\bf 96}, no. 6, 063513 (2017)
  [arXiv:1702.02787 [gr-qc]].
  %%CITATION = doi:10.1103/PhysRevD.96.063513;%%
  %2 citations counted in INSPIRE as of 25 Sep 2017

%\cite{Tu:2017jen}
\bibitem{Tu:2017jen}
  F.~Q.~Tu, Y.~X.~Chen, B.~Sun and Y.~C.~Yang,
  ``Accelerated Expansion of the Universe based on Emergence of Space and Thermodynamics of the Horizon,''
  arXiv:1707.06461 [gr-qc].
  %%CITATION = ARXIV:1707.06461;%%
  %\cite{Cai:2012ip}

\bibitem{Cai:2012ip}
  R.~G.~Cai,
  ``Emergence of Space and Spacetime Dynamics of Friedmann-Robertson-Walker Universe,''
  JHEP {\bf 1211}, 016 (2012)
 % doi:10.1007/JHEP11(2012)016
  [arXiv:1207.0622 [gr-qc]].
  \end{thebibliography}
\end{document}